\documentstyle[twocolumn,prl,aps]{revtex}

\newcommand{\be}{\begin{equation}}
\newcommand{\ee}{\end{equation}}
\newcommand{\lb}{\label}

\newcommand{\oz}{{\overline{z}}}
\newcommand{\bh}{{\bf h}}

\newcommand{\bM}{{\bf M}}
\newcommand{\bV}{{\bf V}}
\newcommand{\bzed}{{\bf 0}}
\newcommand{\balpha}{{\mbox{\boldmath $\alpha$}}}
\newcommand{\bdot}{{\mbox{\boldmath $\cdot$}}}

\begin{document}

\relax

\draft

\title{Turbulence Fluctuations and New Universal Realizability Conditions in
Modelling}
\author{\\
Gregory L. Eyink\\{\em Department of Mathematics}\\
{\em University of Arizona}\\{\em Tucson, AZ 85721}\\
and\\
Francis J. Alexander\\{\em Center for Computational Science}\\
{\em 3 Cummington Street}\\{\em Boston University}\\{\em Boston,
MA 02215}}
\date{\today}
\maketitle

\begin{abstract}
General turbulent mean statistics are shown to be characterized by a
variational principle. The variational
functionals, or ``effective actions'', have experimental consequences for
turbulence fluctuations and are subject
to realizability conditions of positivity and convexity. An efficient
Rayleigh-Ritz algorithm is available to
calculate approximate effective actions within PDF closures. Examples are given
for Navier-Stokes and for a
3-mode system of Lorenz. The new realizability conditions succeed at detecting
{\em a priori} the poor predictions
of PDF closures even when the classical 2nd-order moment realizability
conditions are satisfied.
\end{abstract}
\pacs{PACS Numbers: 47.27.-i, 47.27.Sd, 47.11.+j}

\narrowtext

It does not seem to be a well-recognized fact that general turbulence mean
statistics---such as mean velocity or
pressure profiles---are characterized by a variational principle. However, in
nonequilibrium statistical mechanics
it was pointed out long ago by Onsager \cite{Ons31,OM} that mean histories
satisfy a ``principle of least action''.
The so-called Onsager-Machlup action determines the probability of fluctuations
away from the most probable state.
Close to thermal equilibrium there is a standard fluctuation-dissipation
relation, so that the action has the physical
interpretation of a ``dissipation function''. Onsager's variational principle
reduces then to a principle of least
dissipation.

Recently it has been pointed out by one of us \cite{Ey96I,Ey96II} that a
similar {\em effective action} $\Gamma[z]$
exists in turbulent flow for {\em any} random variable $z(t)$. This action
function {\em (i)} is non-negative ,
$\Gamma[z]\geq 0$, {\em (ii)} has the ensemble mean $\oz(t)$ as its unique
minimum $\Gamma[\oz]=0$, and {\em (iii)}
is convex, $\Gamma[\lambda z_1+(1-\lambda)z_2]\geq
\lambda\Gamma[z_1]+(1-\lambda)\Gamma[z_2],\,\,0<\lambda<1$. These
are realizability conditions which arise from positivity of the underlying
statistical distributions. As a consequence,
the mean value $\oz(t)$ is characterized by a ``principle of least effective
action''. Just as Onsager's action,
this functional is related to fluctuations. In particular, in statistically
stationary turbulence, the time-extensive
limit of the effective action, $V[z]\equiv \lim_{T\rightarrow
\infty}{{1}\over{T}}\Gamma[\{z(t)=z:0<t<T\}]$, the
so-called {\em effective potential}, determines the probability of fluctuations
in the empirical time-average
$\oz_T\equiv {{1}\over{T}}\int_0^T dt\,z(t)$ away from the (time-independent)
ensemble-mean value $\oz$. More
precisely, the probability for any value $z$ of the time-average $\oz_T$ to
occur is given by
\be {\rm Prob}\left(\{\oz_T\approx z\}\right)\sim \exp\left(-T\cdot
V[z]\right). \lb{LD} \ee
This agrees with the standard ergodic hypothesis, according to which, as
$T\rightarrow\infty$, the empirical time-average
must converge to the ensemble-mean, $\oz_T\rightarrow \oz$, with probability
one in every flow realization. The
Eq.(\ref{LD}) refines that hypothesis, by giving an exponentially small
estimate of the probability at a large (but
finite) $T$ to observe fluctuations away from the ensemble-mean.

The realizability conditions on the effective action or effective potential
hold even when there are
no classical 2nd-moment realizability conditions on the means themselves. Thus,
energy spectra or Reynolds
stresses (2nd moments) must be positive, but mean velocity profiles (1st
moments) or energy transfer (3rd moments)
do not satisfy such simple realizability conditions. The new realizability
conditions thus have great potential
significance in modelling if they can be efficiently calculated within
turbulence closures. In \cite{Ey96I,Ey96II} we
have shown that there is a simple Rayleigh-Ritz algorithm within PDF
closures---such as mapping closures \cite{CCK,GK}
or generalized Langevin models \cite{HP,Pope93}---by which the corresponding
approximate values of the effective action
or effective potential may be readily calculated.

As a simple example, we consider first a 3-mode system of Lorenz \cite{Lor60},
in which the equations of motion are
\be \dot{x}_i=A_{i}x_jx_k-\nu_ix_i+f_i, \lb{3mode} \ee
with $i,j,k$ a cyclic permutation of $1,2,3$, with $A_1+A_2+A_3=0$ imposed on
interaction coefficients $A_i$ for energy conservation, with $\nu_i$ positive
damping coefficients, and $f_i(t)$
white-noise random forces with covariance $2\kappa_i$. This dyamics has been
used often as a first test of closure
ideas \cite{Kr63}-\cite{OrsBis}. We consider a simple mapping closure proposed
by Bayly for the 3-mode system \cite{Bay},
which models the realizations by a quadratic map $X_i=\beta_i N_i+\beta_4
N_j'N_k'$ of independent standard Gaussian
variables $N_i,N_i',\,\,i=1,2,3$. The resulting closure equations for the 2nd
moments $M_i=\langle x_i^2\rangle,\,\,i=1,2,3$
and the 3rd moment $T=\langle x_1 x_2 x_3\rangle$ are
\be \dot{M}_i= 2A_i T- 2\nu_i M_i+2\kappa_i \lb{QN123} \ee
for $i=1,2,3$ and
\begin{eqnarray}
\dot{T} & = & A_1M_2M_3+A_2M_1M_3+A_3M_1M_2  \cr
     \, &   & \,\,\,\,\,\,\,\,\,\,\,\,\,\,\,\,\,\,\,\,-(\nu_1+\nu_2+\nu_3)T.
\lb{QN4}
\end{eqnarray}
These are just the {\em quasinormal (QN) equations} for the 3-mode system,
obtained by neglecting the 4th-order
cumulants \cite{PR}. It was already noted by Kraichnan \cite{Kr63} that, unlike
for Navier-Stokes, the QN closure
for the 3-mode system predicts all positive energies. In fact, for
$A_1=2,A_2=A_3=-1,\kappa_1=1,\kappa_2=\kappa_3=0.001,
\nu_1=0.001,\nu_2=\nu_3=1$ it gives steady-state values
\be
\begin{array}{ll}
       \, & M_1^{(QN)}\approx
1.49875,\,\,\,M_2^{(QN)}=M_3^{(QN)}\approx0.50025, \cr
       \,& \,\,\,\,\,\,\,\,\,\,\,\,\,\,\,\,\,\,\,\,T^{(QN)}\approx -0.49925.
\end{array}
\lb{QNval} \ee
All of the 2nd moments are positive, as required by realizability. However, DNS
of the 3-mode dynamics itself gives
\be
\begin{array}{l}
      M_1^{(DNS)}=4.46\pm 0.03\cr
      M_2^{(DNS)}=M_3^{(QN)}=0.49876\pm 0.00002\cr
      T^{(DNS)}= -0.49776\pm 0.00002.
\end{array}
\lb{DNSval} \ee
While the QN predictions for $M_2,M_3$ and $T$ are within ${{1}\over{3}}\%$ of
the DNS values, $M_1$ is underpredicted
by $66\%$ in the QN approximation. As is well-known, satisfaction of
realizability cannot guarantee that a prediction is
correct. However, failure of realizability certainly implies that the
predictions are in error. In Figs.1-3 we graph the
approximate effective potentials of the energy variables
$E_i={{1}\over{2}}x_i^2$ and triple product $T=x_1x_2x_3$ in the
$QN$ closure as calculated by the Rayleigh-Ritz algorithm for the above PDF
model. The numerical method is outlined below
and described in detail in \cite{Ey96II,AlEy96}. It is apparent that $V_{E_2}$
and $V_T$ satisfy realizability but that
$V_{E_1}$ does not. Thus, one may infer {\em a priori}---without knowledge of
the DNS results---that the QN prediction for
the mean of $E_1$ is not converged. In this case, the failure of realizability
of the predicted $V_{E_1}$ succeeds at
detecting the poor prediction for the mean value, even though the classical
2nd-moment condition $E_1\geq 0$ is satisfied.
In the same plots in Figs.2-3 we have graphed also the effective potentials
$V_{E_2}$ and $V_T$ obtained from DNS. They
do not agree with the QN potentials as closely as do the corresponding means:
the accurate prediction of fluctuations
is a much more stringent demand on the closure. However, we note that the
predictions of Bayly's PDF closure are at least
qualitatively correct for $V_{E_2}$ and $V_T$ and give correctly the order of
magnitude of the averaging-time needed
to eliminate fluctuations in those variables. Of course, the prediction of
$V_{E_1}$ is not even qualitatively correct.

The Rayleigh-Ritz algorithm used in obtaining the approximate potentials from
the PDF closure involves a fixed point problem
very similar to (and, in fact, generalizing) the fixed point condition
determining the predicted steady-state moments
themselves. The system of equations that must be solved in general is
\be {{\partial V_0}\over{\partial \bM}}(\bM,\bh)\alpha_0
+\left({{\partial\bV}\over{\partial\bM}}\right)^\top(\bM,\bh)\bdot\balpha=
V_0(\bM,\bh)\balpha, \lb{FP1} \ee
\be \bV(\bM,\bh)=V_0(\bM,\bh)\bM, \lb{FP2} \ee
\be \alpha_0+\balpha\bdot\bM=1. \lb{FP3} \ee
The vector $\bM=(M_1,...,M_k)$ contains the moments of the closure, e.g., in
the case above, $k=4$ (and $M_4=T$). $\bh$ is
the vector of ``perturbation fields'', one associated with each variable $Z_i$
for which the potential is to be determined.
In our previous calculation $\bh=(h_{E_1},h_{E_2},h_T)$. When $\bh=\bzed$ the
vector $\bV(\bM,\bh)$ coincides with the
dynamical vector $\bV(\bM)$ which appears in the closure equation:
$\dot{\bM}=\bV(\bM)$ (cf. Eqs.(\ref{QN123}),(\ref{QN4})
above). The perturbations for $\bh\neq {\bf 0}$ are determined by the method
discussed in \cite{Ey96II}. The 0-component
$V_0(\bM,\bh)$ is associated with the zeroeth moment $M_0\equiv 1$ and it may
be written explicitly here as $V_0(\bM,\bh)
={{1}\over{2}}h_{E_1}M_1+{{1}\over{2}}h_{E_2}M_2+h_TM_4.$ It is easy to check
that, when $\bh=\bzed$, the stationary moments
$\bM_*$ along with $\alpha_{*0}=1,\balpha_*=\bzed$ solve the system
Eqs.(\ref{FP1})-(\ref{FP3}). Once the solutions
$\alpha_{*0}(\bh),\balpha_*(\bh),\bM_*(\bh)$ are known for $\bh\neq\bzed$, the
effective potential $V_{Z_i}$
is constructed as a function of $h_i$ via $V_{Z_i}[h_i]=
-\balpha_{*}(h_i)\bdot\bV\left(\bM_*(h_i)\right)$. To obtain the
potential as a function of $z_i$, the expected value ${\bf Z}_*(\bh)={\bf z}$
must be inverted to give $h_i$ as
a function of $z_i$. For full details of the algorithm, see
\cite{Ey96II,AlEy96}.

Our results point toward significant new directions in turbulence modelling.
The new realizability conditions apply
individually to {\em all} predicted means. We see above that they can
successfully discriminate between poor predictions
for one set of variables and good predictions for another. Calculating each
point on the graph of an effective potential
curve within a closure requires just the same amount of computation as that to
calculate the predicted mean. It is
therefore very easy to apply the above realizability conditions as a check to
detect poor predictions in advance, without
expensive testing by experiment or simulation. This gives a strong incentive to
the development of PDF closures, such as
those in Refs.\cite{CCK}-\cite{Pope93}. In conjunction with our variational
method they can give some {\em a priori}
information in turbulence modelling. This is a unique advantage, almost never
obtained in other statistical closure methods.

It remains to be seen how well the new realizability conditions succeed in
detecting poor predictions of closures
for Navier-Stokes turbulence. It is thus worthwhile to give one example
demonstrating the Rayleigh-Ritz method for a
statistically time-dependent Navier-Stokes flow. The simplest such situation is
freely-decaying homogeneous and isotropic
turbulence with random initial data. We consider a model energy spectrum
\be E(k,t)= \left\{ \begin{array}{ll}
                     Ak^m   & k\leq k_L(t) \cr
                     \alpha\varepsilon^{2/3}(t)k^{-5/3} & k_L(t)\leq k \leq
k_d(t) \cr
                     0      & k\geq k_d(t)
                     \end{array} \right.
\lb{spectrum} \ee
which has been adopted before in this problem \cite{C-BC,Rey}. As long as
$0<m<4$ it is commonly believed that
there is a permanence of the low-wavenumber spectrum. This motivates one to
adopt the above self-preserving
form, in which the shape of the spectrum is unchanged in time except through
its dependence on the parameters $\varepsilon(t),
k_L(t)$ and $k_d(t)$. At high Reynolds number there is only one independent
such parameter, since the relation $k_L(t)=
\left({{\alpha}\over{A}}\varepsilon(t)\right)^{{{3}\over{3m+5}}}$ is required
by continuity and, when $k_L(t)\ll k_d(t)$,
$k_d(t)=\left({{4}\over{3\alpha\nu}}\right)^{3/4}\varepsilon^{1/4}(t)$ also
holds \cite{Rey}. The remaining time-dependence
is determined by considering the evolution of the mean energy
$E(t)={{1}\over{2}}\langle v^2(t)\rangle$.
For the above form of the spectrum it is not hard to show \cite{Rey} that the
dissipation $\varepsilon(t)
={{\nu}\over{2}}\sum_{ij}\langle (\partial_i v_j+\partial_j v_i)^2\rangle$ is
given as
\be \varepsilon(t)= \Lambda_m\cdot E^p(t) \lb{diss} \ee
with
$\Lambda_m^{-1}=\alpha^{3/2}\left({{1}\over{m+1}}
+{{3}\over{2}}\right)^{{{3m+5}\over{2m+2}}}A^{{{1}\over{m+1}}}$ and
$p={{3m+5}\over{2m+2}}$. Thus, employing the Navier-Stokes equation via its
energy-balance, one obtains the closed equation
\be \dot{E}(t)= -\Lambda_m\cdot E^p(t). \lb{energ-eq} \ee
Its solution gives a prediction for the energy-decay law, as $E_*(t)\sim
(t-t_0)^{-{{2m+2}\over{m+3}}}$: see \cite{Rey}.

It is interesting to make a check on the various hypotheses involved in these
predictions by means of the effective
action $\Gamma[E]$ for the energy history $E(t)$. As a simple PDF model for the
above closure, one may adopt a
Gaussian random velocity field with the assumed self-similar spectrum
Eq.(\ref{spectrum}). The Rayleigh-Ritz approximation
of the effective action within the Gaussian ansatz can be analytically
evaluated \cite{Ey96III}, with the result:
\begin{eqnarray}
\, & & \Gamma^{(Gauss)}[E]={{3}\over{2(p-2)\Lambda_m}}\times \cr
\, & & \int_0^\infty dt \,\,{{\left(\dot{E}(t)+\Lambda_m\cdot K^p(t)\right)
                    \left(\dot{K}(t)+\Lambda_m\cdot
K^p(t)\right)}\over{K^{p+1}(t)}}
\lb{Gauss-eff}
\end{eqnarray}
where $K(t)$ is a variational parameter satisfying
\begin{eqnarray}
\, & & \Lambda_m\cdot K^p(t)+\dot{E}(t)= \cr
\, & & \,\,\,\,\,\,\,\,\,\,\,\,\,\,\,(p-2)\Lambda_m\cdot(E(t)-K(t))\cdot
K^{p-1}(t). \lb{var-parm}
\end{eqnarray}
It is easy to check that, if the predicted closure mean energy $E_*(t)$
satisfying $\dot{E}_*(t)= -\Lambda_m E^p_*(t)$
is substituted, then $\Gamma^{(Gauss)}[E_*]=0$. Further insight is obtained by
considering small perturbations $E(t)=
E_*(t)+\delta E(t)$ from the predicted mean. By a straightforward calculation
it follows that
\begin{eqnarray}
\, & & \Gamma^{(Gauss)}[E]={{3}\over{8(p-1)\Lambda_m}}\times \cr
\, & & \int_0^\infty dt\,\,{{\left(\delta\dot{E}(t)+\Lambda_m\cdot
pE^{p-1}_*(t)\delta E(t)\right)^2}
                \over{E^{p+1}_*(t)}} + O(\delta E^3).
\lb{quad-eff}
\end{eqnarray}
This is the same law of fluctuations as would be realized with the Langevin
equation
\be \delta\dot{E}(t)+\Lambda_m\cdot pE^{p-1}_*(t)\delta E(t)=
\sqrt{2R_*(t)}\eta(t) \lb{Lang-eq} \ee
obtained by linearization of the energy-decay equation around its solution
$E_*(t)$ and by addition of a white-noise
random force $\eta(t)$ with strength
\be R_*(t)={{2(p-1)}\over{3}}\varepsilon_*(t)E_*(t). \lb{FDT} \ee
Thus, the smaller fluctuations from the ensemble-mean value are predicted to
decay according to a linearized law, similar to
the Onsager regression hypothesis for equilibrium fluctuations. Likewise, the
expression Eq.(\ref{FDT}) is a
{\em fluctuation-dissipation relation} analogous to that in equilibrium. These
are testable predictions of the Gaussian
closure. Note that the coefficient $(p-1)$ in front of the action is $>0$ as
long as $m>-3$. In fact, $m>-1$ is required to
give a finite energy. Thus, for all permissable values of $m$, the approximate
action  $\Gamma^{(Gauss)}[E]$ satisfies
realizability. One should be cautioned again that satisfaction of realizability
is only a consistency check and cannot
guarantee correctness of predictions. Failure of realizability, as observed in
the 3-mode model, is more
practically useful, although in a purely negative way.

The previous examples and our variational method are discussed in greater
detail in forthcoming papers
\cite{AlEy96,Ey96III}. Here, we have simply wished to illustrate briefly the
use of the action principle. Future work will
study the success of the new realizability conditions in detecting poor closure
predictions for more realistic Navier-Stokes
flows, of greater interest to practical engineering. It should be clear that
very general PDF ansatz may be employed in our
method, either by guessing a functional form of the PDF or by hypothesizing
``surrogate'' random variables to model the
actual flow realizations. Any guess of the turbulence statistics---such as the
``synthetic turbulence'' models of
\cite{JLSS}---may be input to yield predictions for realistic problems. We
therefore expect our method to be a flexible
framework within which to develop novel turbulence closures. Insights from
simulation, experiment and recent theoretical
developments can be readily incorporated. The advantage of the variational
formulation is that it provides built-in checks
of statistical closures which may detect a sizable fraction of faulty
predictions in advance. By doing so cheaply, it can
provide great savings in turbulence modelling for practical engineering
purposes.

\noindent {\bf Acknowledgements:} We wish to thank R. H. Kraichnan for his
interest in and encouragement of
this work. Numerical computations were carried out at the Center for
Computational Science at Boston University
and the Department of Mathematics at the University of Arizona.

\newpage
\begin{center}
FIGURE CAPTIONS
\end{center}

Figure 1.)
Effective potential for energy in mode 1
in quasinormal closure.

Figure 2.)
Effective potential for energy in mode 2
in quasinormal closure. (DNS with errorbars).

Figure 3.)
Effective potential for triple moment
in quasinormal closure. (DNS with errorbars).


\begin{references}

\bibitem{Ons31}L. Onsager, Phys. Rev. {\bf 37} 405 (1931); {\bf 38} 2265
(1931).
\bibitem{OM}L. Onsager and S. Machlup, Phys. Rev. {\bf 91} 1505 (1953).
\bibitem{Ey96I}G. L. Eyink, J. Stat. Phys. {\bf 83} 955 (1996).
\bibitem{Ey96II}G. L. Eyink, Phys. Rev. E, to appear (1996), {\em
chao-dyn/9505001}.
\bibitem{CCK}H. Chen, S. Chen and R. H. Kraichnan, Phys. Rev. Lett. {\bf 65}
575 (1990).
\bibitem{GK}T. Gotoh and R. H. Kraichnan, Phys. Fluids A {\bf 5} 445 (1992).
\bibitem{HP}D. C. Haworth and S. B. Pope, Phys. Fluids {\bf 29} 387 (1986).
\bibitem{Pope93}S. B. Pope, Ann. Rev. Fluid Mech. {\bf 26} 23 (1994).
\bibitem{Lor60}E. Lorenz, Tellus {\bf 12} 243 (1960).
\bibitem{Kr63}R. H. Kraichnan, Phys. Fluids {\bf 6} 1603 (1963).
\bibitem{Kr67}R. H. Kraichnan, in: {\em Dynamics of Fluids and Plasmas}, ed. S.
I. Pai
                  (Academic Press, New York, 1966), pp.239-255.
\bibitem{OrsBis}S. A. Orszag and L. R. Bissonnette, Phys. Fluids {\bf 10} 2603
(1967).
\bibitem{Bay}B. Bayly, unpublished (1992).
\bibitem{PR}I. Proudman and W. H. Reid, Phil. Trans. Roy. Soc. A {\bf 247} 163
(1954).
\bibitem{AlEy96}G. L. Eyink and F. J. Alexander, in preparation.
\bibitem{C-BC}G. Comte-Bellot and S. Corrsin, J. Fluid Mech. {\bf 25} 657
(1966).
\bibitem{Rey}W. C. Reynolds, Ann. Rev. Fluid Mech. {\bf 8} 183 (1976).
\bibitem{Ey96III}G. L. Eyink, in preparation (1996).
\bibitem{JLSS}A. Juneja et al. Phys. Rev. E {\bf 49} 5179 (1994).

\end{references}
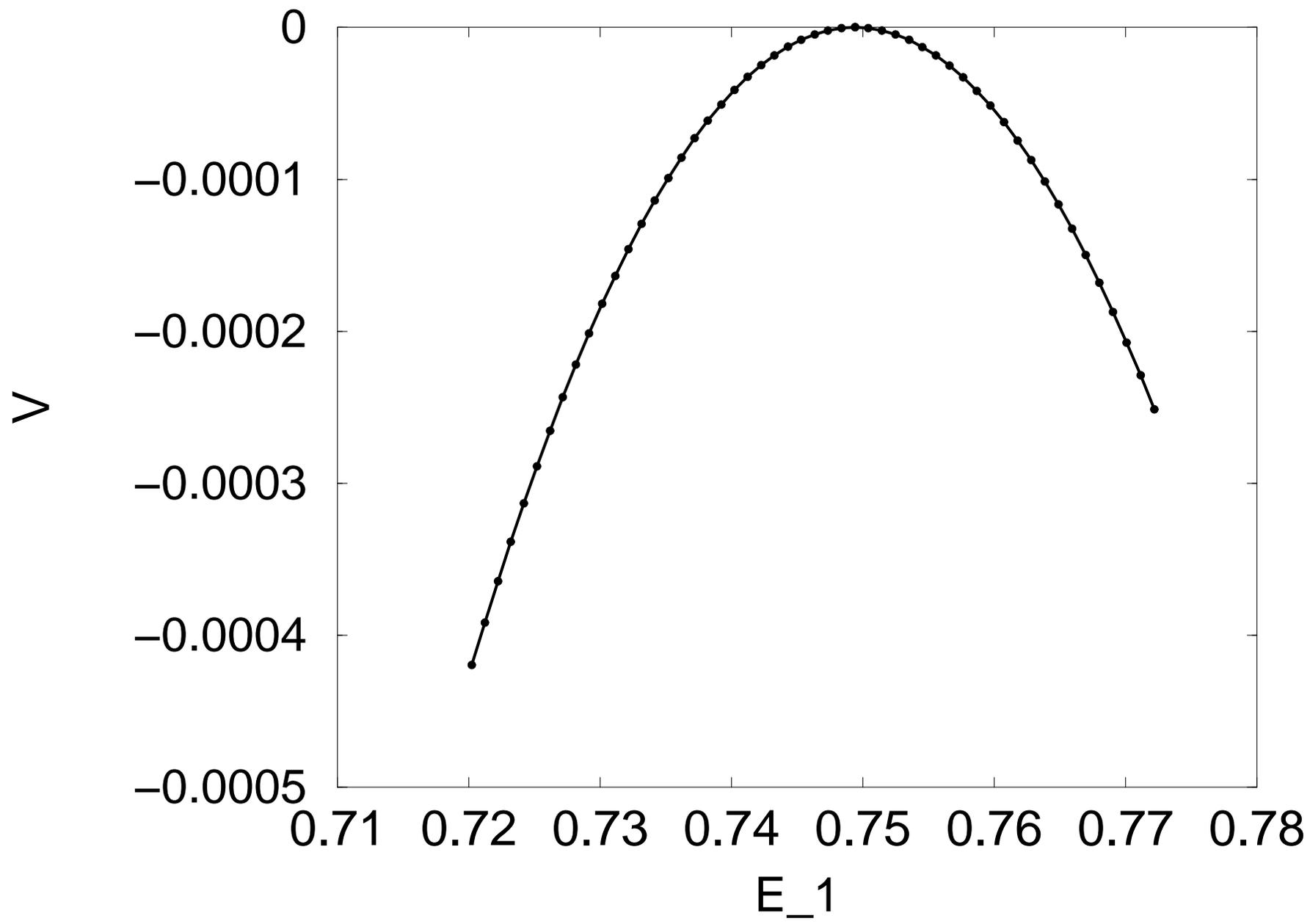
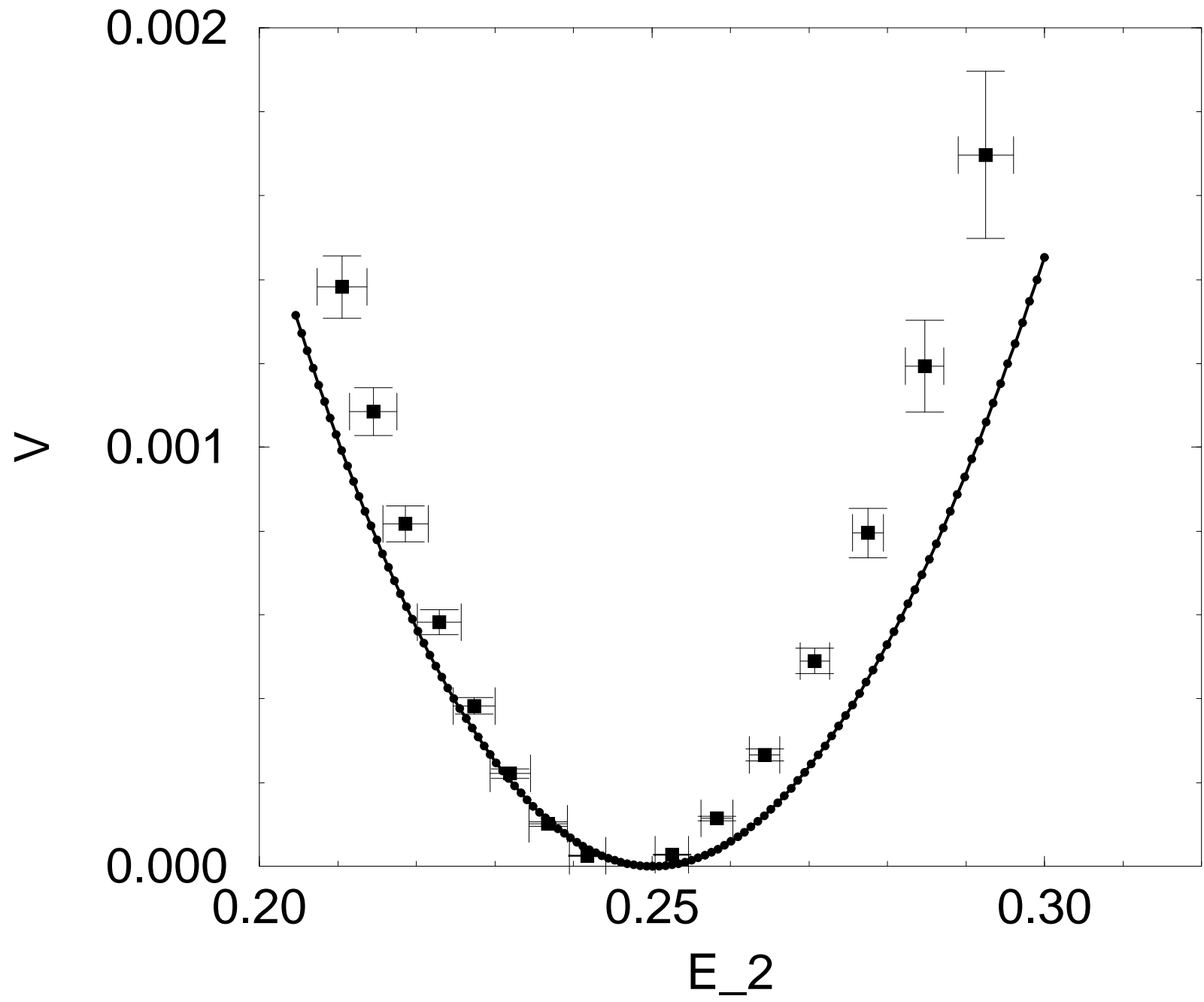
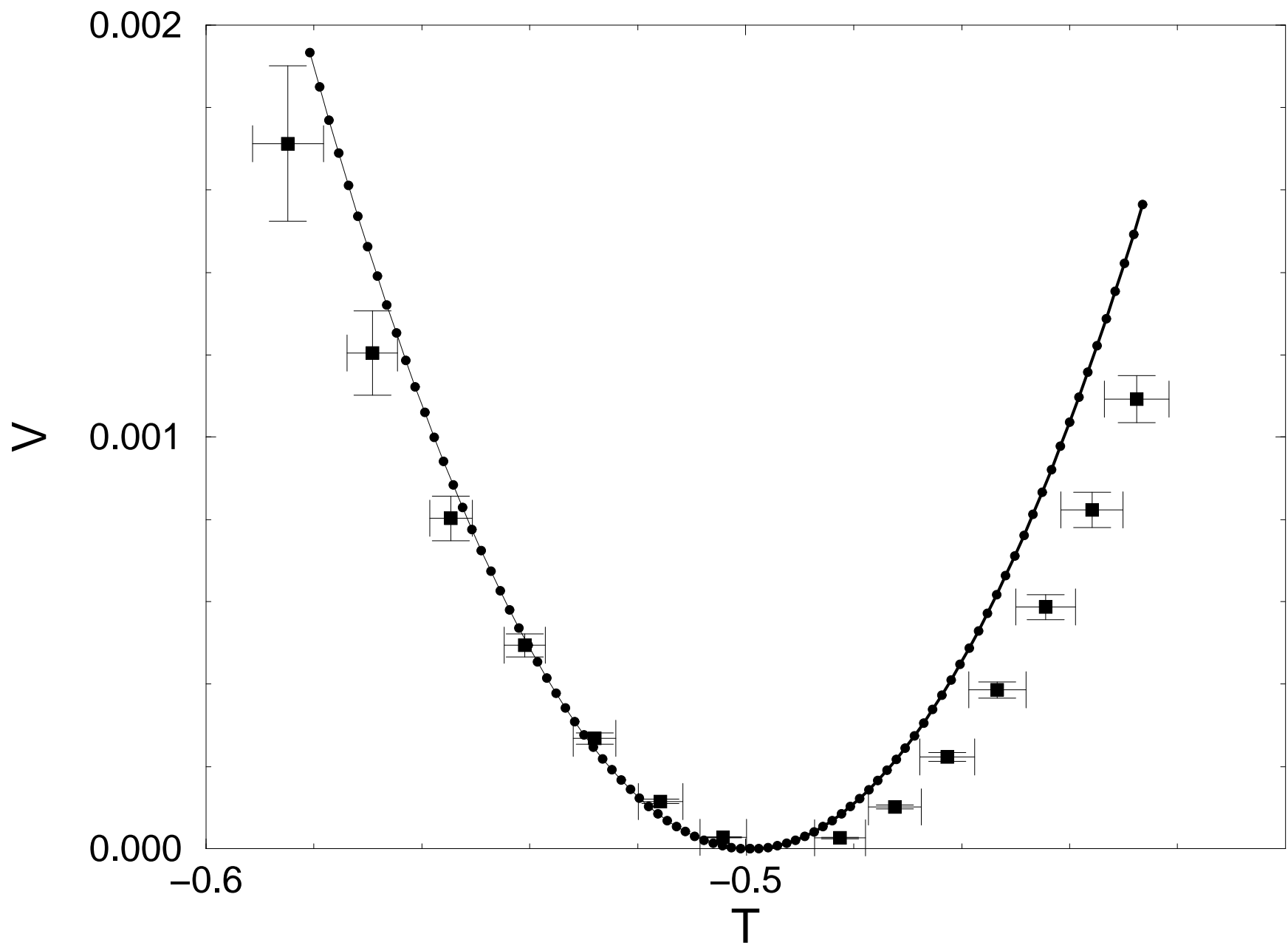
\end{document}